\documentclass[]{spie}  %>>> use for US letter paper
\usepackage{amsmath,amsfonts,amssymb}
\usepackage{graphicx}
\usepackage[colorlinks=true, allcolors=blue]{hyperref}
\usepackage{xcolor}
\usepackage{gensymb}
\usepackage{booktabs}
\usepackage{bm}
\usepackage{array}

\newif\ifdraft
\drafttrue

\definecolor{orange}{rgb}{1,0.5,0}
\definecolor{gr}{rgb}{0,0.65,0}
\definecolor{mygray}{gray}{0.95}

\ifdraft
 \newcommand{\RS}[1]{{\color{red}{\bf RS: #1}}}
 
 \newcommand{\PMN}[1]{{\color{orange}{\bf PMN: #1}}}

\else
 \renewcommand{\sout}[1]{}
 \newcommand{\RS}[1]{{\color{red}{}}}
 
 \newcommand{\PMN}[1]{{\color{red}{}}}
 
\fi

\title{Inferring geometry and material properties from Mueller matrices with machine learning}

\author[a,b]{Lars Doorenbos}
\author[a]{C.~H.~Lucas Patty}
\author[a]{Raphael Sznitman}
\author[a]{Pablo Márquez-Neila}
\affil[a]{University of Bern, Bern, Switzerland}
\affil[b]{University of Bonn, Bonn, Germany}

\authorinfo{Further author information: (Send correspondence to L.D.)\\L.D.: E-mail: doorenbos@iai.uni-bonn.de}

\begin{document}

\maketitle

\begin{abstract}
Mueller matrices (MMs) encode information on geometry and material properties, but recovering both simultaneously is an ill-posed problem. We explore whether MMs contain sufficient information to infer surface geometry and material properties with machine learning. 
We use a dataset of spheres of various isotropic materials, with MMs captured over the full angular domain at five visible wavelengths (450-650 nm). 
We train machine learning models to predict material properties and surface normals using only these MMs as input. 
We demonstrate that, even when the material type is unknown, surface normals can be predicted and object geometry reconstructed.
Moreover, MMs allow models to identify material types correctly.
% , and we learn a self-supervised material embedding that maps MMs of unseen materials into informative clusters, enabling the characterization of novel materials.
Further analyses show that diagonal elements are key for material characterization, and off-diagonal elements are decisive for normal estimation.
%Finally, we show the benefits of using full MMs over measurements from unpolarized light only (i.e., the first MM column). 
\end{abstract}

%
% Uncomment for keywords
\vspace{2pc}

\section{Introduction}

The ability to accurately retrieve geometric and material properties of surfaces through optical characterization techniques is of paramount importance across a broad range of scientific and technological domains. These methods have demonstrated significant utility in diverse applications, including but not limited to remote sensing~\cite{tyo2006review}, where surface features must be inferred from aerial or satellite imagery, and in (bio-)medical imaging~\cite{he2021polarisation}, where optical responses can reveal critical information about tissue structure and composition. Despite this promise, achieving precise and reliable estimations of such properties remains a complex and unresolved challenge. %However, precise determination of these properties remains non-trivial due to limitations in our understanding of the precise relations, along with practical factors such as processing time and noise. These issues are further exacerbated when estimating multiple properties simultaneously.

Out of the available optical characterization approaches, polarimetry is a powerful technique for analyzing the optical properties of surfaces. In particular, Mueller matrices (MMs) are known to encode information on both geometry and material properties. Despite this, inferring these characteristics directly from the MMs remains challenging and often requires strong assumptions about either the material or the geometry of the measured object.
In this work, we adopt a data-driven approach to overcome the reliance on these assumptions and use machine learning~(ML) to assess to which extent MMs contain sufficient information to recover these properties. 

We validate this by presenting a straightforward ML model capable of predicting surface normals directly from a single MM, without requiring any additional information. Notably, this prediction remains feasible, to an extent, when the material type was not present in the training data. We show this experimentally by generalizing the estimation of the model to previously unseen materials. Furthermore, we show that ML models can also leverage MMs to identify what material type it stems from, regardless of the orientation of the surface from which the measurement was taken. We conclude with additional analyses that confirm the necessity for full MM polarimetry and provide practical guidelines that enable users to more efficiently design their detectors when only subsets of properties are of interest.

\section{Methods}

\subsection{Data}

We use the publicly available dataset from Baek et al. \cite{baek2020image}, which consists of complete $4 \times 4$ MMs acquired for 25 different isotropic spherical materials. The dataset contains measurements across the full angular domain, obtained at five wavelengths in the visible spectrum (450, 500, 550, 600, and 650\,nm). The materials in the dataset exhibit a wide variety of characteristics, including diffuse and specular surfaces, metallic and dielectric surface compositions, varying surface roughness and different spectral albedos. We normalize each MM by its top-left element $\text{MM}_{1,1}$.

\subsection{Machine learning algorithm}

We use machine learning models to predict material properties and surface normals from the MMs. We consider these problems separately. Both cases are examples of a supervised learning problem, where the ML algorithm learns the mapping from the inputs (i.e., the MMs) to the desired output properties. 

More specifically, for normal estimation, we learn the mapping from MMs $M \in \mathcal{M}$ to the corresponding normal vectors with a function (model) $f: \mathcal{M} \to \mathcal{R}^3$, while for material classification the output comes in the form of a label $y \in Y$, where the model $g: \mathcal{M} \to Y$ assigns a material to the input MM.

We have the choice between many supervised machine learning methods to use for $f$ and $g$. We use the well-established random forest~\cite{breiman2001random} (RF) algorithm for our experiments. Our choice for RFs is motivated by their speed, accuracy, and interpretability in the form of feature importances.
We use the RF implementation of \texttt{scikit-learn}~\cite{pedregosa2011scikit}.

We now provide a more detailed discussion on the experimental setting for the two tasks.

\subsection{Normal estimation}

Our first experiments investigate the possibility of predicting the surface normals from their corresponding Mueller matrices. We randomly split the MMs pixel-wise into training, validation, and testing splits, and $f$~is trained on the training set only. We design three experimental settings to evaluate the performance of the model: 
\begin{description}
    \item[Single-material training.] We train and test a separate model for each of the 25~materials individually. This allows us to assess the model performance when there is no material variation.
    \item[All-material training.] We train a single model using MMs from all 25~materials combined. This setup evaluates the model's ability to generalize across different material types when they are represented during training.
    \item[Leave-one-material-out (LOMO) generalization.] We conduct 25~experiments, each time excluding a material from training and testing on it. This setup evaluates the model's capacity to generalize to previously unseen materials.
\end{description}

% . In the first, only a single material is considered at a time for a total of 25 experiments. In the second, we use all 25 materials simultaneously. In the third, we perform another 25 experiments, where we leave out one material of the training set, and report the results of evaluating the learned model on the test set of the left-out material. This allows us to probe the extrapolation capacities of the RF.

Building on the leave-one-material-out experiment, we further investigate how the model's ability to generalize improves as more materials are progressively included in the training set. In this setup, we fix a set of eight test materials and start by training the RF on a single material. We then incrementally add one material at a time to the training set and evaluate performance on the same fixed test set after each addition. This process continues until all 25 materials are included. To assess the impact of material ordering, we repeat the experiment with two different permutations of the training material sequence.

In all cases, we measure the accuracy of the normal estimation by computing the average angular error (in radians) between the predicted and ground-truth normals on the test set.

Finally, we test whether the predicted normals can be used to reconstruct the shape of the object from which the MMs were captured. To this end, we predict the normals of the testing split, which were not seen by the RF during training. We then interpolate the predicted normals to obtain a dense normal map covering every pixel. The surface shape is recovered from these normals by first computing a gradient field and then integrating it to produce the final height map. We follow the depth-from-normals approach of~\cite{Merker_Depth_from_Normals_2022}.

\subsection{Material classification}

The second experiments explore the extent to which we can predict the material type from the MMs. 

In this case, we split the spheres into two halves. The top half is used as the testing split. The bottom half is randomly divided into training and validation sets, at a ratio of 80:20. Again, $g$ is trained on the training set only. All materials are used simultaneously. As such, the task is to recognize to which material (out of 25) an MM taken at an unseen normal belongs. 

We use accuracy to measure the quality of the material classification, which is defined as the ratio of correct predictions:
\begin{equation}
    \text{Accuracy} = \frac{|\text{Correct predictions}|}{|\text{All predictions}|}.
\end{equation}

% \subsection{Unseen material clustering}

% We aim to extract a representation from the Mueller matrices that clusters unseen materials into representative groups. To do so, we use supervised contrastive learning, which optimizes a neural network with the supervised contrastive loss given by \cite{khosla2020supervised}
% \begin{equation}
% \mathcal{L}_{con} = \sum_{i\in I}\frac{-1}{|P(i)|} \sum_{j\in P(i)}\log \frac{\exp(\bm{z}_i \cdot \bm{z}_j / \tau)}{\sum_{k\in A(i)} \exp(\bm{z}_i \cdot \bm{z}_k / \tau)},
% \end{equation}
% where $\bm{z}$~represents normalized latent representations of a sample, and $\tau$~the temperature. $A(i) \equiv I \setminus \{i\}$ denotes all other samples in the batch, and $P(i) \equiv \{p \in A(i) : y_p = y_i\}$ gives all samples in the batch with the same label as $i$.

\section{Results and discussion}

\subsection{Normal estimation}

\begin{figure*}[t]
  \centering
  \setlength\tabcolsep{3pt}
  \begin{tabular}{ccc}
    \includegraphics[width=0.3\linewidth]{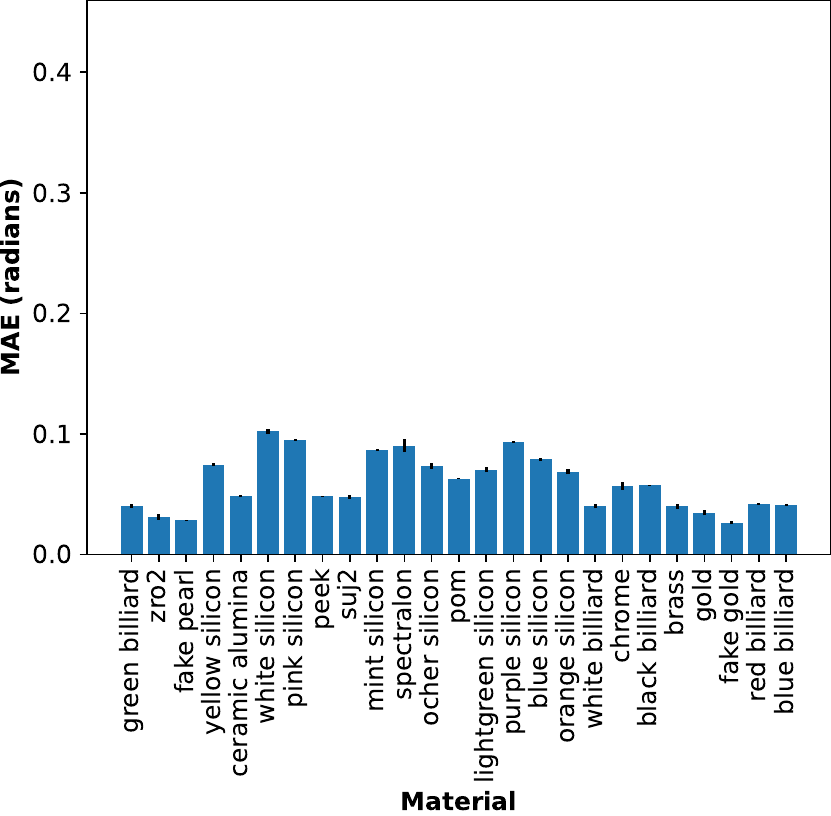} &
    \includegraphics[width=0.3\linewidth]{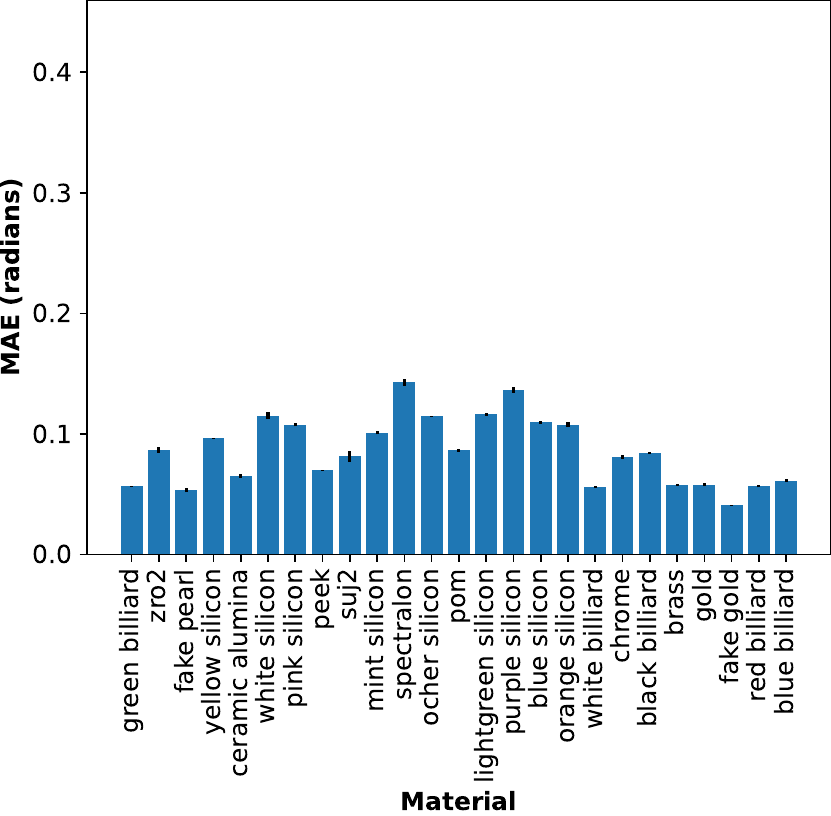} &
    \includegraphics[width=0.3\linewidth]{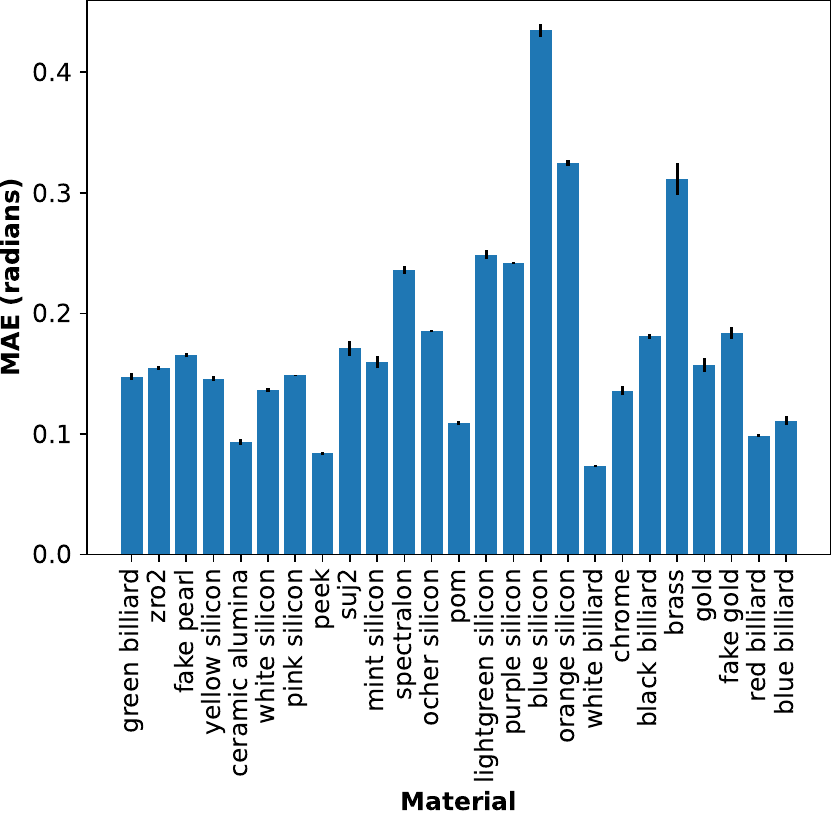} \\
  \end{tabular}
    \caption{\textbf{Normal estimation performance with (a)~a single material at a time, (b)~all materials together, and (c)~leave-one-material-out.} We show the mean and standard deviation of the angular error (in radians) over three runs. }
    \label{fig:norm_est}
\end{figure*}

We show results for the normal estimation experiments in Fig.~\ref{fig:norm_est}. We find that the RF can predict the unseen normals from their Mueller matrices with high accuracy in both Fig.~\ref{fig:norm_est}(a), where each bar is obtained by training and testing on that individual material, and Fig.~\ref{fig:norm_est}(b), where all materials are trained and tested on simultaneously. Nonetheless, the specialized models of Fig.~\ref{fig:norm_est}(a) that train and test on single materials achieve better results than the single, general model of Fig.~\ref{fig:norm_est}(b). While the overall trend in relative performance on the materials is similar, there are still some differences in results between the two experiments. For instance, the error on zro2 more than doubles in the second experiment, performing worse than green billiard. In both cases, fake pearl and fake gold show the lowest error. In contrast, the worst-performing materials are white and pink silicon for the single-material experiment and purple silicon and spectralon for the experiment with all materials simultaneously.

Our leave-one-material-out experiment (Fig.~\ref{fig:norm_est}(c)) shows that the model can predict normals for unseen materials to a significant extent, although performance is, as expected, lower than in the previous two settings. The poorest performance is observed for blue silicon, with an average angular error of 0.45~radians. The model also struggles to generalize to orange silicon and brass.

% In Fig.~\ref{fig:norm_est}(c), we show results of our leave-one-material-out experiment for extrapolating the normal prediction to unseen materials. In most cases, the model can predict the normals to a large extent, although the results are unsurprisingly worse than in the previous two experiments. Blue silicon is the worst-performing material, where the average prediction is 0.45~rads off. The model also struggles to extrapolate to orange silicon and brass.

% \subsubsection{Extrapolation}

\begin{figure}[t]
  \centering
  \begin{tabular}{cc}
    \includegraphics[width=0.45\linewidth]{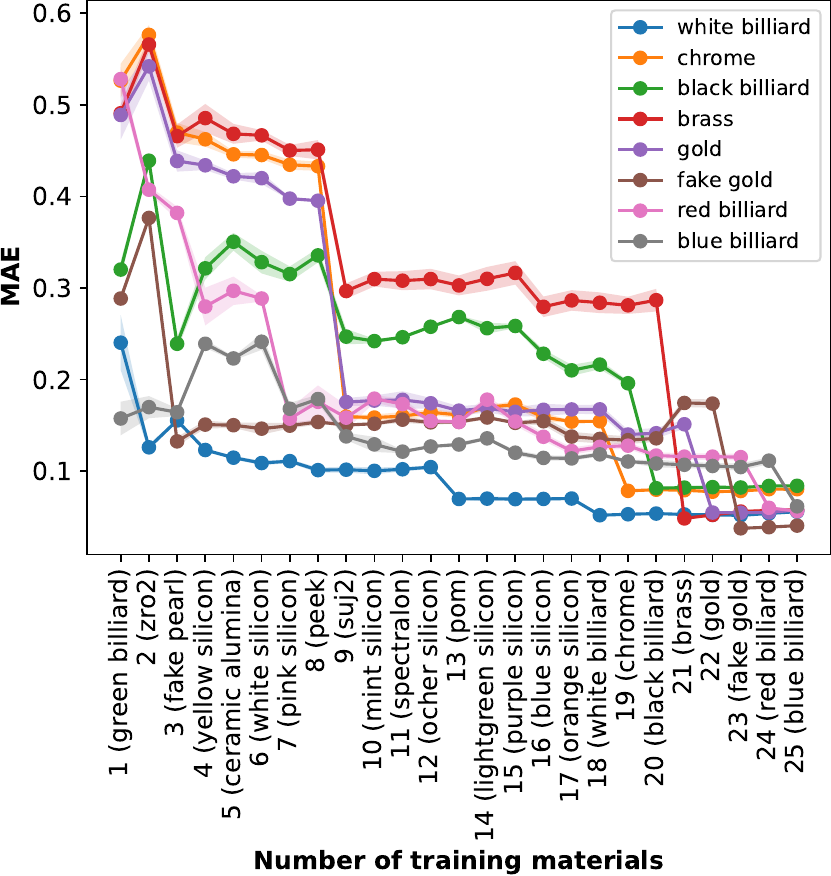} &
    \includegraphics[width=0.45\linewidth]{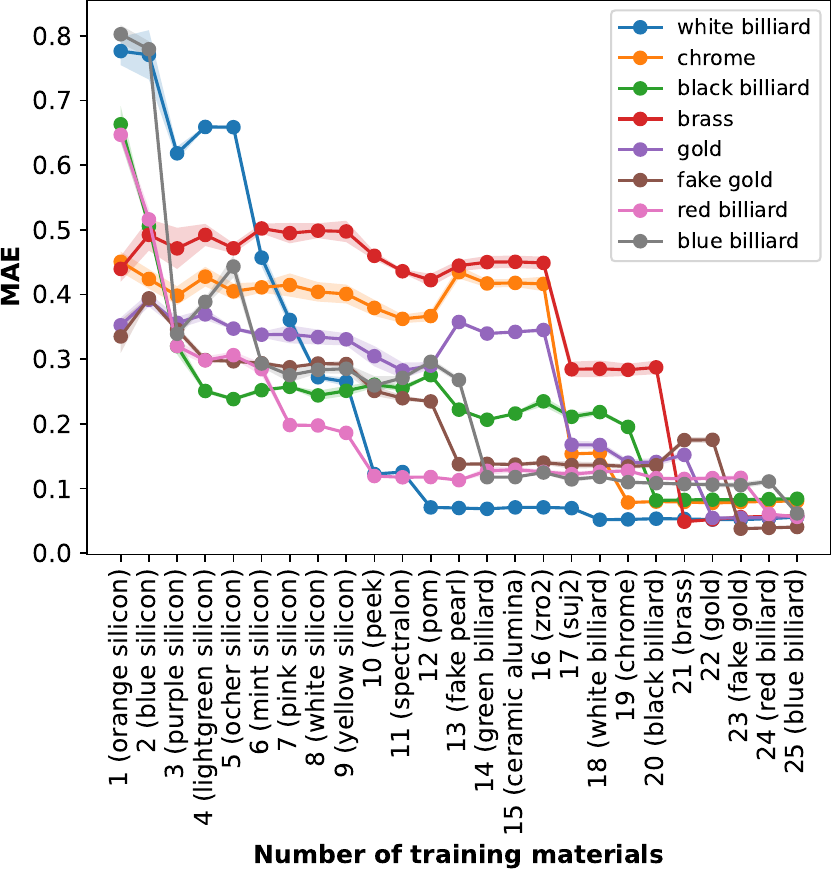} \\
    (a) & (b) \\
  \end{tabular}
    \caption{\textbf{Extrapolating normal prediction to unseen materials.} We progressively add more materials to the training set to see how the performance on new materials evolves over time. (a) and (b) show two different orderings of the training materials added. The test materials are the same for both. }
    \label{fig:extra}
\end{figure}

Fig.~\ref{fig:extra} shows the results of progressively adding materials to the training set and tracking the performance on eight fixed test materials. As expected, the performance improves significantly when the test material itself is eventually added to the training set. For instance, the error for brass decreases from 0.3 to approximately 0.05 radians when brass is included in training (Fig.~\ref{fig:extra}(a)).

Interestingly, certain materials yield substantial performance improvements for others. For example, including fake pearl leads to a nearly threefold improvement in the performance on fake gold, and including the metal suj2 improves results for both gold and brass. In general, while adding more training materials improves generalization to unseen materials, we observe notable performance jumps when materials with similar characteristics are added.

Furthermore, repeating the experiment with different insertion orders shows that the points of performance gains vary considerably, further suggesting that similarity between materials plays a key role in the model's generalization behavior.

\subsection{Shape-from-ellipsometry}

\begin{figure}[t]
  \centering
  \begin{tabular}{cc}
    \includegraphics[width=0.45\linewidth]{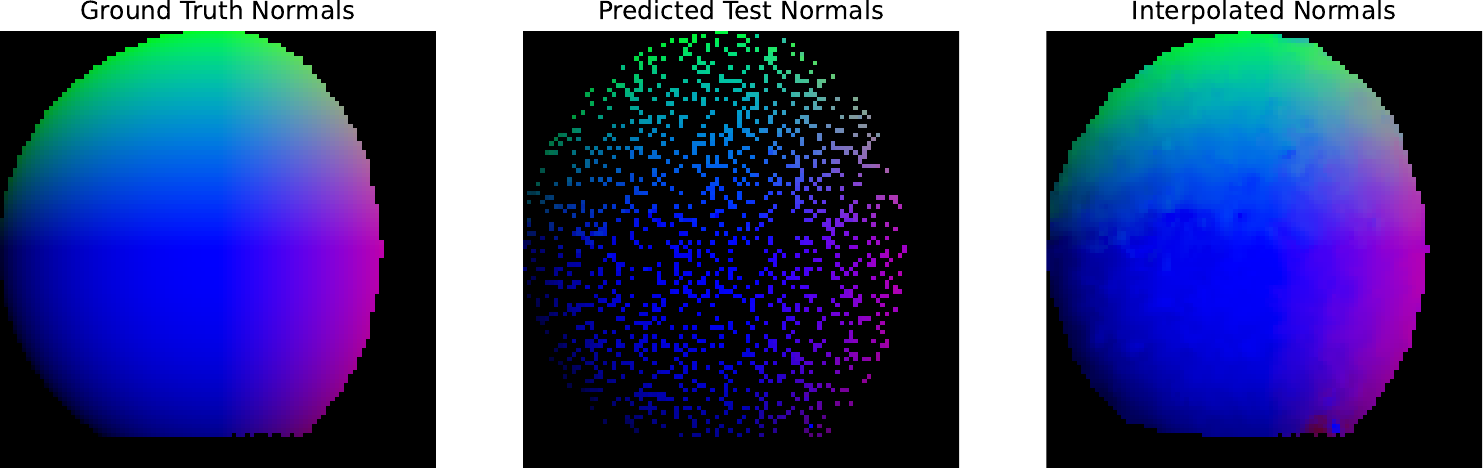} &    \includegraphics[width=0.45\linewidth]{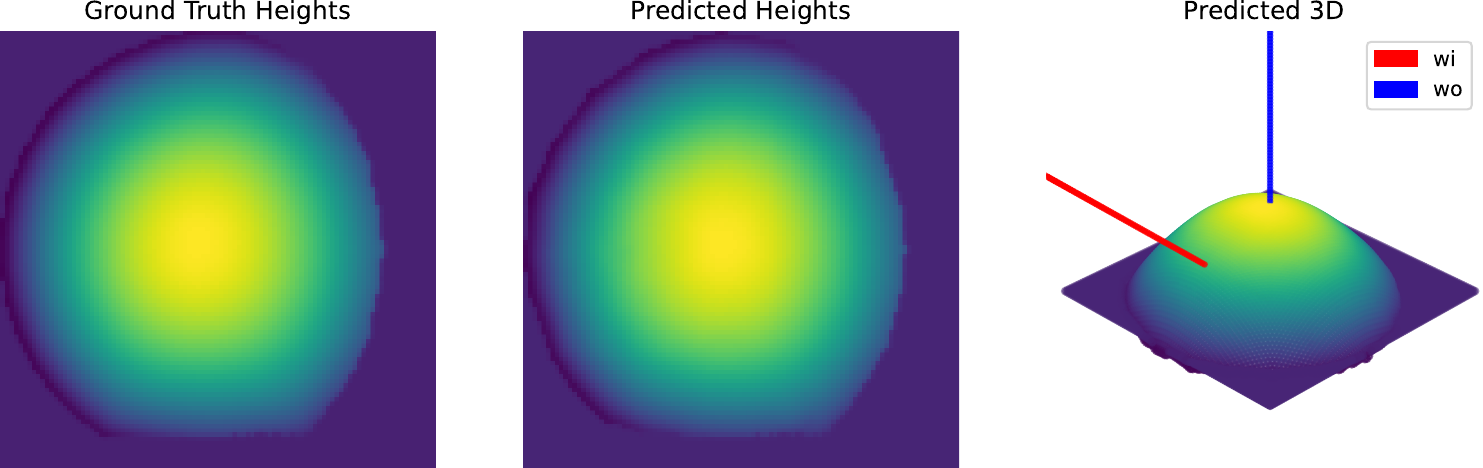} \\
    \includegraphics[width=0.45\linewidth]{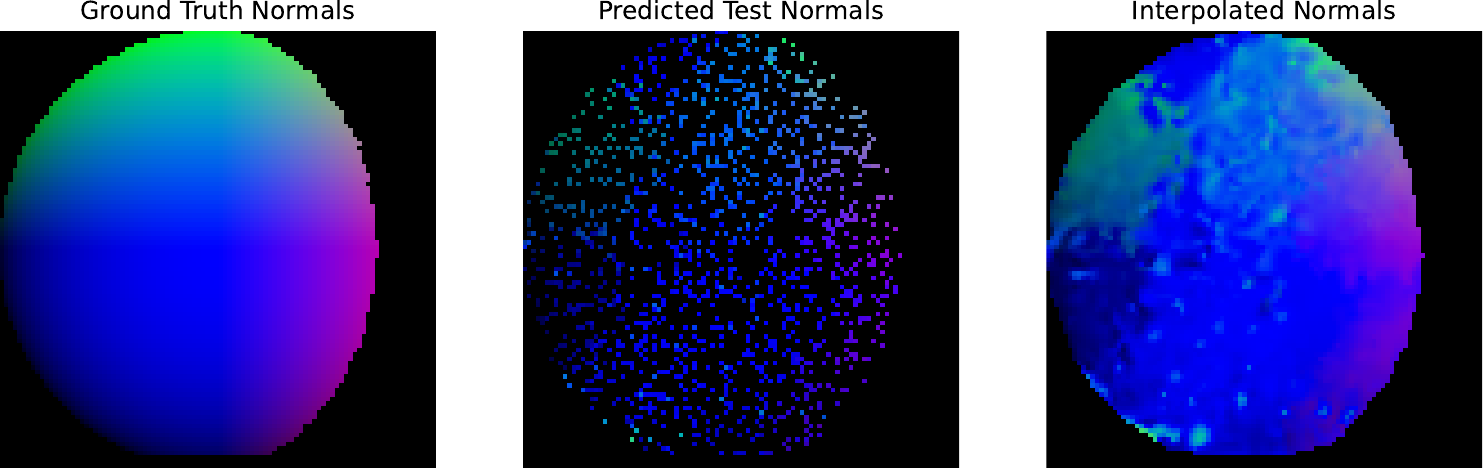} &    \includegraphics[width=0.45\linewidth]{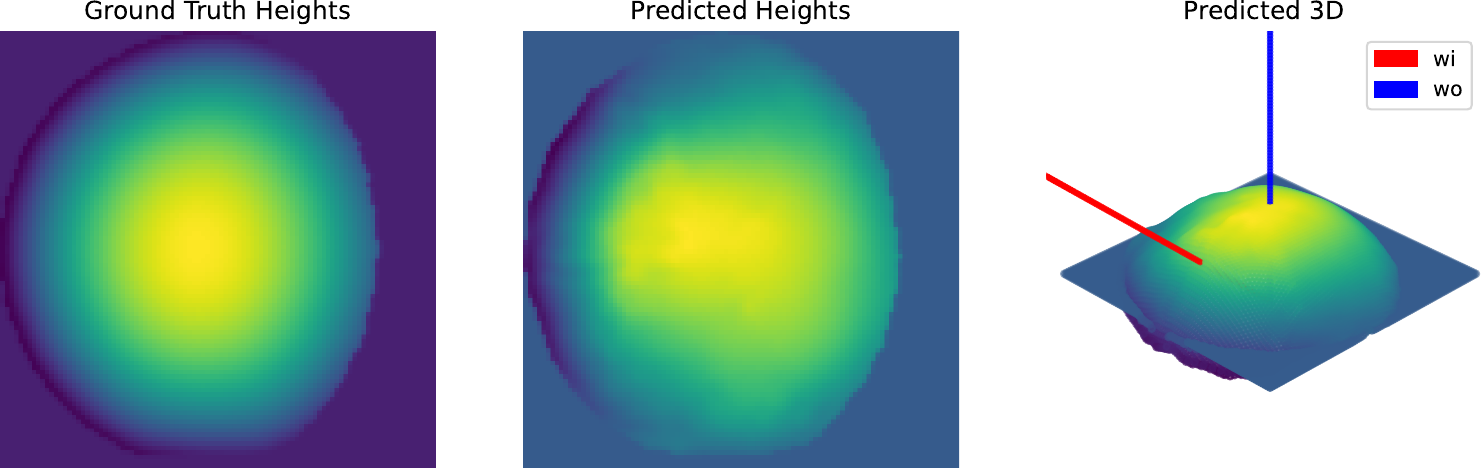} \\
    \includegraphics[width=0.45\linewidth]{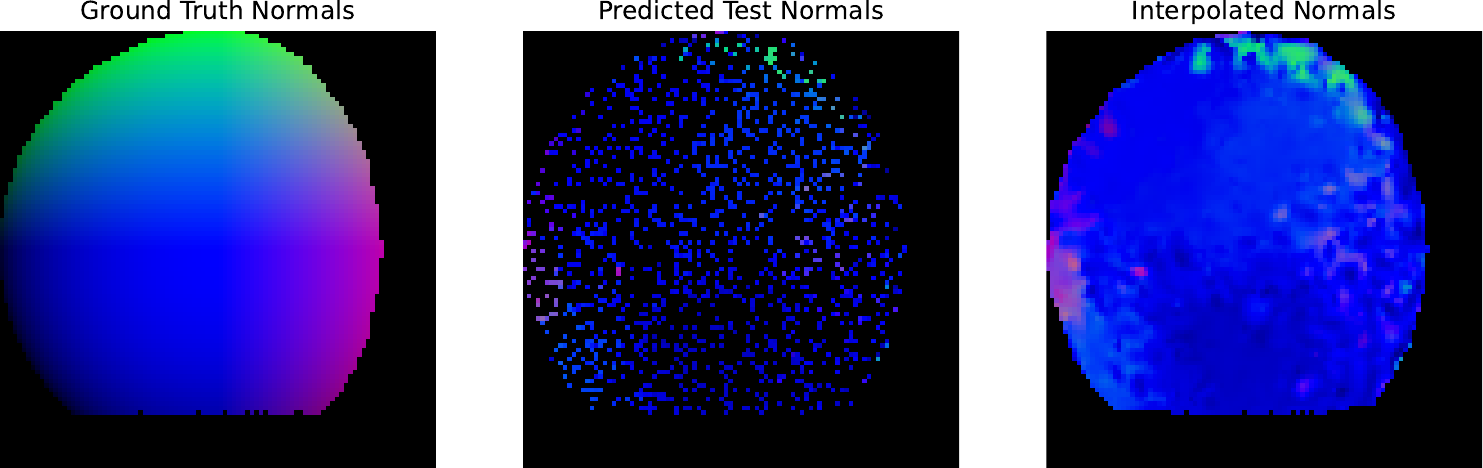} &    \includegraphics[width=0.45\linewidth]{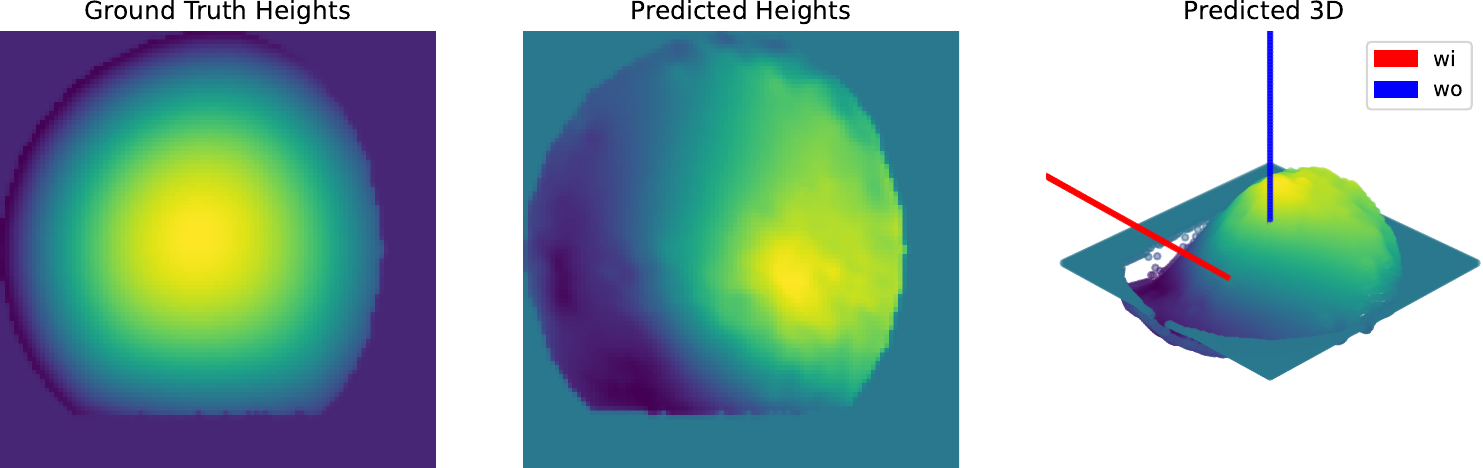} \\
  \end{tabular}
    \caption{\textbf{Recovering the shape from predicted normals.} The top row shows the results for training and testing on green billiard. The middle row shows results for training on green billiard, and testing on the unseen material white billiard. The bottom row was trained on green billiard and tested on chrome. We can reconstruct the underlying shape of the sphere from predictions on unseen Mueller matrices, even when the material is different but related. }
    \label{fig:sfe}
\end{figure}

We successfully reconstruct the shape of the test object by integrating the height map derived from the interpolated test normals predicted on the \emph{green billiard} material (Fig.~\ref{fig:sfe}(top)). The result closely matches the original spherical shape, even though the model had never seen these normals during training.

% In Fig.~\ref{fig:sfe}(top), we show the reconstruction of the underlying shape of the object through integration of the height map from the interpolated test normals on green billiard. We can successfully obtain the sphere that was measured. Notably, this reconstruction comes only from predictions on Mueller matrices from unseen normals.

To assess cross-material generalization, we test the model reconstructions on different materials. When predicting normals on a \emph{white billiard} sphere, the reconstruction remains largely accurate (Fig.~\ref{fig:sfe}(middle)). In contrast, performance degrades when using \emph{chrome} as the test material (Fig.~\ref{fig:sfe}(bottom)), suggesting that material similarity also plays a key role in reconstruction quality.

% Additionally, we include experiments where the test material is different from the train material. In the middle row, we show that a white billiard sphere is still reconstructed to a large extent, whereas the bottom row shows that chrome is too different from green billiard to successfully do so.

\subsection{Material characterization}

\begin{figure}[t]
  \centering
  \setlength\tabcolsep{3pt}
  \begin{tabular}{cc}
    \includegraphics[width=0.45\linewidth]{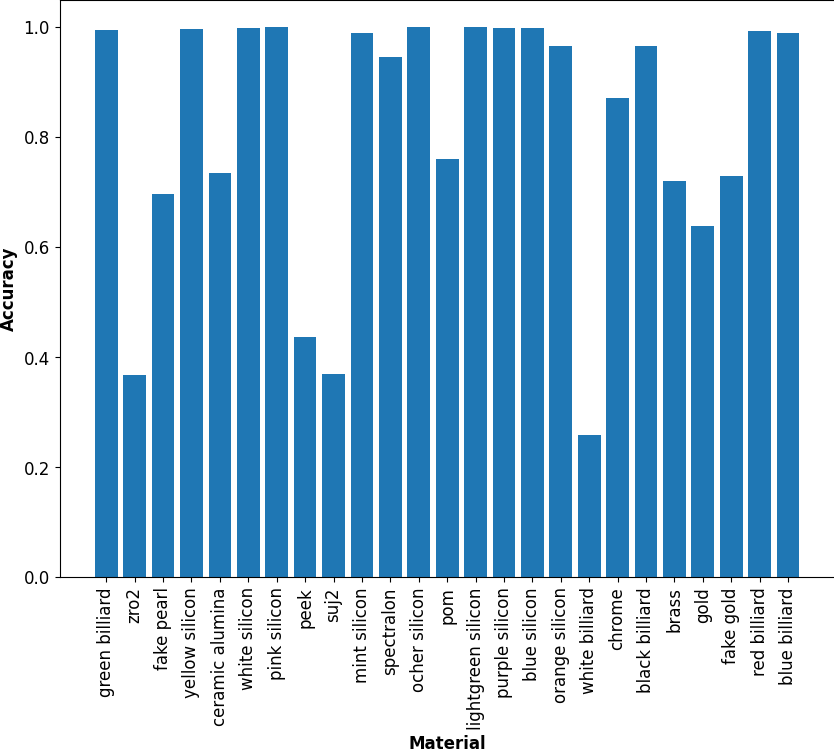} &
    \includegraphics[width=0.45\linewidth]{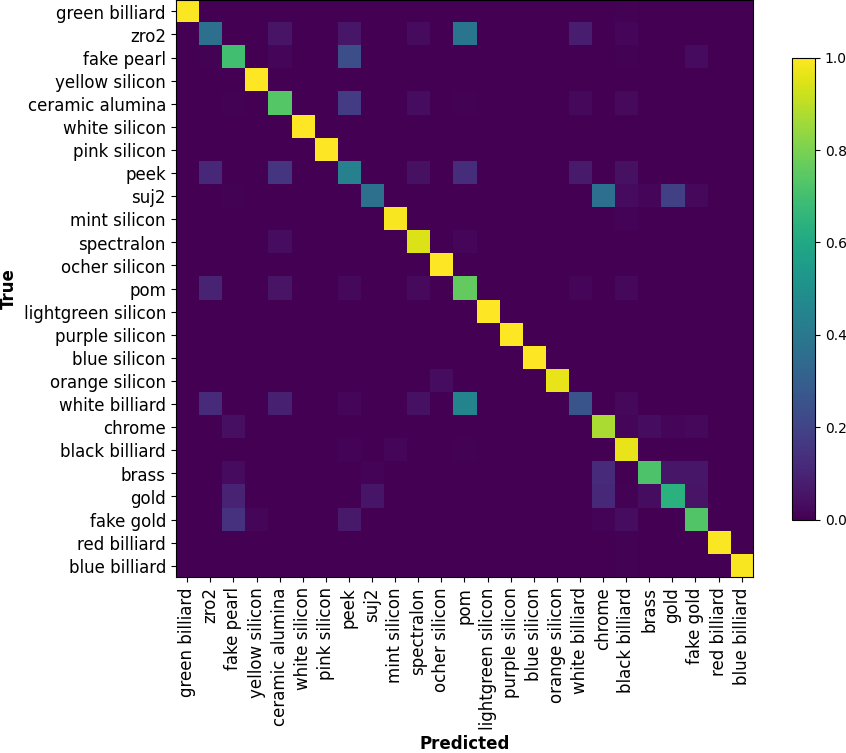} \\
  \end{tabular}
    \caption{\textbf{Classification performance per material and the confusion matrix of the RF for material classification.} The RF can distinguish the materials on unseen geometry with a high accuracy of~81\%. }
    \label{fig:mat_class}
\end{figure}

We evaluate the random forest performance on the 25-class material classification task, with results shown in Fig.~\ref{fig:mat_class}. The model achieves near-perfect accuracy (close to~1.0) on many of the silicone and billiard materials, with the exception of \emph{white billiard}, which is the most difficult material to classify in the dataset. The confusion matrix shows that this is due to predictions confusing MMs from \emph{white billiard} with those of the material \emph{pom}.

% We show results with an RF for the 25-way material classification problem in Fig.~\ref{fig:mat_class}. The accuracy on many of the silicone and billiard materials is close to perfect~($1.0$), with the exception of white billiard, which is the most difficult to recognize material in the entire dataset. The confusion matrix shows that this is due to predicting MMs measured from white silicon \PMN{meant \emph{white billiard}?} as the material \emph{pom}.

Performance on other materials is more varied. For example, accuracy on \emph{zro2}, \emph{peek}, and \emph{suj2} is around 40\%, while materials such as \emph{ceramic alumina}, \emph{brass}, and \emph{gold} fall in the 60-80\% range. The confusion matrix suggests that these lower scores are due to similarities between certain materials. For instance, \emph{gold} is often mistaken for other visually or structurally similar materials, such as \emph{chrome}, \emph{fake gold}, or \emph{fake pearl}, which likely exhibit similar properties.

\subsection{Feature importance}

We measure the contribution of each MM element to the model's predictions by computing their feature importances for the two tasks considered. Higher importance values indicate that an element plays a greater role in the model's decision-making process.

% Our choice for RF allows us to identify which elements of the MMs are the most important for the different tasks, by looking at the feature importances. These measure the contribution of individual elements to the overall decision made by the model. Higher values indicate a more important element for the given task.

\begin{figure}
    \centering
    \includegraphics[width=\linewidth]{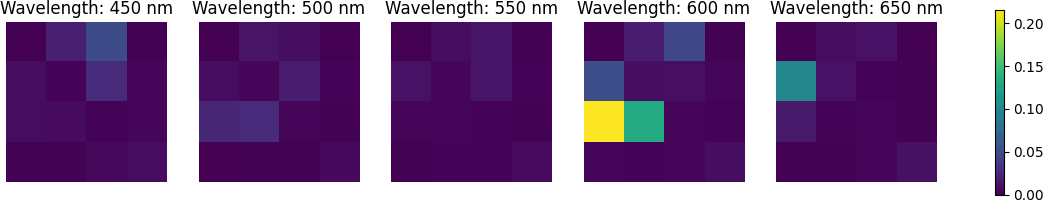}
    \caption{\textbf{Feature importances for the normal estimation task.} The linear off-diagonal elements of the Mueller matrices are the most informative to find the normals.}
    \label{fig:norm_imp}
\end{figure}

\begin{figure}[t]
  \centering
  \begin{tabular}{c}
    \includegraphics[width=\linewidth]{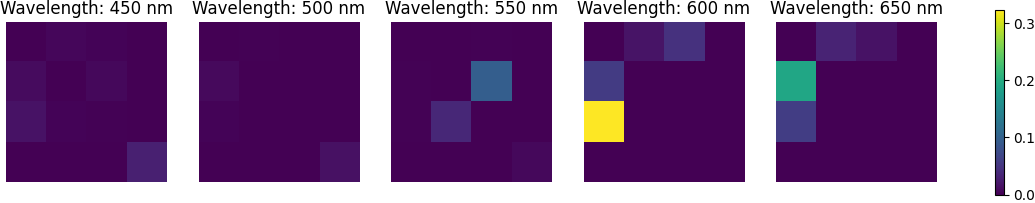} \\
    (a) \\
    \includegraphics[width=\linewidth]{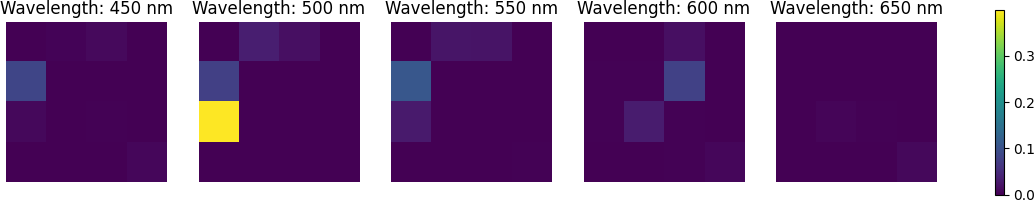} \\
    (b) \\
    \includegraphics[width=\linewidth]{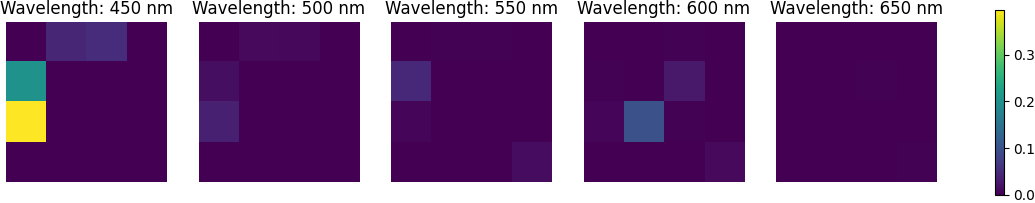} \\
    (c) \\
  \end{tabular}
    \caption{\textbf{Feature importances for the normal estimation task on three individual materials: (a) red, (b) green, and (c) blue billiard.} We see again that the off-diagonal elements of the Mueller matrices are the most informative, but there are clear differences between materials depending on their color. }
    \label{fig:norm_imp_idv}
\end{figure}

\textbf{Normal estimation.} 
Feature importances for the normal estimation task are shown in Fig.~\ref{fig:norm_imp}. We find the off-diagonal elements to be the most important across all wavelengths, especially at 600\,nm. This aligns with the physical intuition that off-diagonal elements encode polarization cross-coupling effects, which are more sensitive to surface orientation and therefore more informative for estimating normals.

To investigate the effect of color, we also show feature importances on three individual materials (red, green, and blue billiard) in Fig.~\ref{fig:norm_imp_idv}. The results indicate that the most important features vary depending on the color of the material.

\begin{figure}
    \centering
    \includegraphics[width=\linewidth]{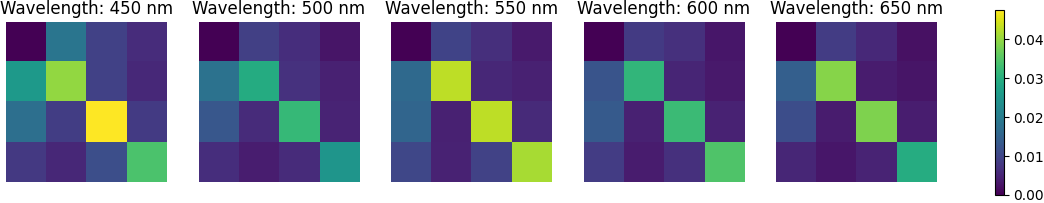}
    \caption{\textbf{Feature importances for the classification task.} The diagonal elements of the Mueller matrices are the most informative to discriminate materials regardless of wavelength.}
    \label{fig:class_imp}
\end{figure}

\textbf{Material characterization.}

Feature importances for the classification task are shown in Fig.~\ref{fig:class_imp}. We find that the diagonal elements are the most important features across all wavelengths. 

This is expected, as diagonal elements reflect how the material attenuates or preserves specific polarization states, which are closely related to intrinsic optical properties such as reflectance, absorption, and scattering. These properties are key factors in material characterization.

\subsection{Mueller matrix feature subsets}

\begin{table}[t]
\caption{\textbf{Results with different subsets of the Mueller matrices.} Going beyond unpolarized incident light, i.e., only the first column, is crucial to reach high performance. }
    \label{tab:sub}
    \centering
    \begin{tabular}{c|cc}
    \toprule
         & Normals & Materials\\
        \midrule
        Full ($d=15$) & 0.085 & 81.5 \\
        Unpolarized ($d=3$) & 0.180 & 50.0 \\        
        % Upper ($d=9$) & 0.168 & 82.0 \\
        Diagonal ($d=3$) & 0.341 & 81.3\\
        Off-diagonal ($d=12$) & 0.101 & 59.9\\
        Linear ($d=8$) & 0.094 & 73.0 \\
        Circular ($d=3$) & 0.362 & 72.7\\
        \bottomrule
    \end{tabular}
\end{table}

The feature importance analysis suggests that off-diagonal elements primarily contribute to normal estimation, while diagonal elements are more relevant for material classification. To investigate this further, we conduct additional experiments in which the RF is trained on specific feature subsets of the MMs, each reflecting different physical properties. The considered feature subsets are:
\begin{description}
    \item[Unpolarized.] The first column of the MM, excluding the top-left element $\text{MM}_{1,1}$, which is always set to~1 after normalization. 
    \item[Upper.] All elements on or above the main diagonal.
    \item[Diagonal.] The three diagonal elements of the MM (excluding $\text{MM}_{1,1}$). 
    \item[Off-diagonal.] All elements not on the main diagonal.
    \item[Linear.] The top-left 3x3 block of the MM, corresponding to the linear polarization components.
    \item[Circular.] The elements in the corners of the MM: $\text{MM}_{1,4}$, $\text{MM}_{4,1}$, and $\text{MM}_{4,4}$.
\end{description}

Tab.~\ref{tab:sub} summarizes the results. Using the off-diagonal elements yields a performance close to that of the full matrix for normal estimation. However, this subset leads to a 21.6~percentage point drop in material classification accuracy. In contrast, using only the diagonal elements results in classification performance comparable to the full matrix, but with normal estimation error increasing by a factor of more than four. These findings align with the feature importance patterns observed in Figs.~\ref{fig:norm_imp},\ref{fig:norm_imp_idv}, and \ref{fig:class_imp}, and further highlight the importance of linear polarization for normal estimation.

A particularly interesting comparison involves the unpolarized subset, using only the first column of the MM. In this case, performance is substantially reduced for both tasks. The average angular error increases from 0.085 to 0.18~radians for normal estimation, and material classification accuracy drops from 81.5\% to 50\%. These results show the benefits of using the full MM.

Finally, the results with linear and circular subsets show that neither alone is sufficient for good performance. Combining both subsets yields the best results, especially for material classification.

\subsection{Effect of the phase angle}

\begin{figure}[t]
  \centering
  \setlength\tabcolsep{3pt}
  \begin{tabular}{cc}
    \includegraphics[width=0.5\linewidth]{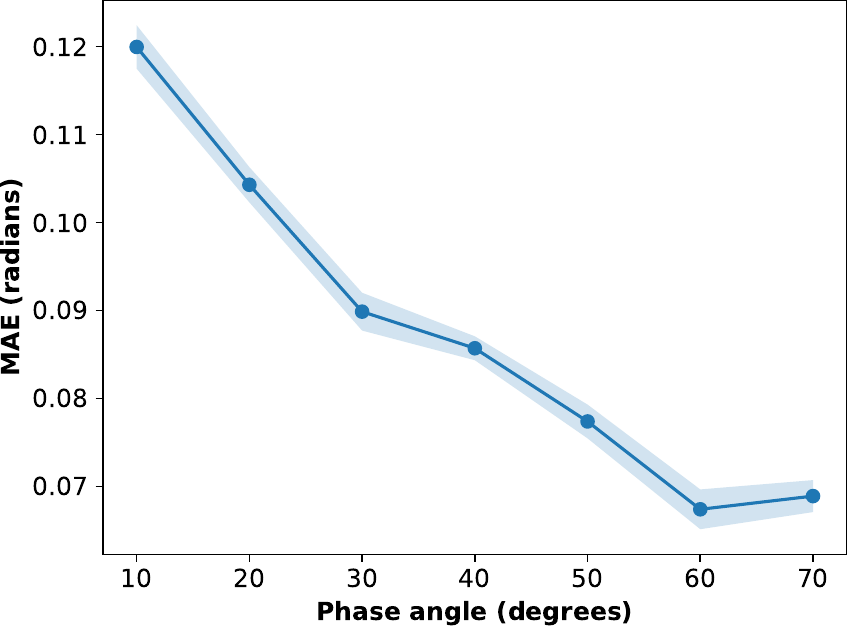} &
    \includegraphics[width=0.5\linewidth]{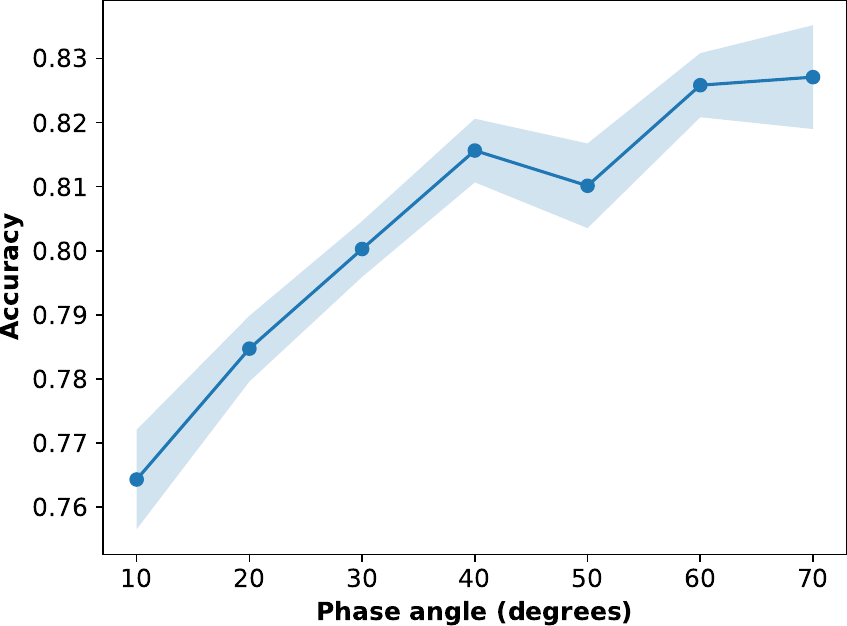} \\
    (a) & (b) \\
  \end{tabular}
    \caption{\textbf{Effect of the phase angle on (a) normal estimation and (b) material classification performance.} We show the mean and standard deviation over three runs. Higher phase angles lead to better results on both tasks.}
    \label{fig:pa}
\end{figure}

While we set the phase angle to 40\degree throughout the previous experiments, the dataset~\cite{baek2020image} allows us to evaluate the RF performance as a function of the phase angle. In this analysis, we train separate RF models for both tasks using data acquired at various phase angles, and report the results in Fig.~\ref{fig:pa}.

A key consideration is that the data for different phase angles contains different regions of the spheres; higher phase angles cover smaller areas. To ensure a fair comparison, we restrict training and testing to the overlapping region visible at all considered phase angles and compare performance within a narrow angular range. While this reduces the amount of usable data, it enables a consistent and balanced evaluation across phase angles.

We find that increasing the phase angle improves performance on both tasks, with the most rapid gains occurring at lower angles. Beyond approximately 50~degrees, the performance begins to plateau. These findings are consistent with our expectations: for material classification, higher phase angles approach the Brewster angle, which is typically used for analyzing material properties. For normal estimation, a larger phase angle increases the contrast in reflectance across different surface orientations, enhancing the model's ability to distinguish between them.

\section{Conclusions}

We have shown that machine learning can effectively recover information about both material type and surface orientation from a Mueller matrix. Our models accurately predict surface normals and object geometry across a range of experimental conditions, enabling downstream analyses such as surface shape reconstruction.
Moreover, the machine learning models achieve high accuracy in classifying different materials. A key finding is that using the full MMs significantly outperforms approaches relying only on unpolarized incident light. While the full MM is important for optimal performance on both tasks, our analysis also reveals that specific subsets of MM elements can be used to reduce acquisition time and complexity when only a subset of the properties is of interest. Together, these results offer practical guidance for designing future polarimetric systems and highlight the potential of data-driven methods in optical surface and material characterization.

% Together, these findings provide new insights for practitioners looking to improve surface estimation or material classification results through optical characterization.

\acknowledgments
This work was funded by the Swiss National Science Foundation (SNSF), research grant 200021\_192285 “Image data validation for AI systems”.

% \clearpage

% \appendix
% \input{05appendix}

\bibliographystyle{spiebib}
\bibliography{refs} 

\begin{thebibliography}{1}

\bibitem{tyo2006review}
Tyo, J.~S., Goldstein, D.~L., Chenault, D.~B., and Shaw, J.~A., ``Review of passive imaging polarimetry for remote sensing applications,'' {\em Applied optics}~{\bf 45}(22),  5453--5469 (2006).

\bibitem{he2021polarisation}
He, C., He, H., Chang, J., Chen, B., Ma, H., and Booth, M.~J., ``Polarisation optics for biomedical and clinical applications: a review,'' {\em Light: Science \& Applications}~{\bf 10}(1),  194 (2021).

\bibitem{baek2020image}
Baek, S.-H., Zeltner, T., Ku, H., Hwang, I., Tong, X., Jakob, W., and Kim, M.~H., ``Image-based acquisition and modeling of polarimetric reflectance.,'' {\em ACM Trans. Graph.}~{\bf 39}(4),  139 (2020).

\bibitem{breiman2001random}
Breiman, L., ``Random forests,'' {\em Machine learning}~{\bf 45},  5--32 (2001).

\bibitem{pedregosa2011scikit}
Pedregosa, F., ``Scikit-learn: Machine learning in python fabian,'' {\em Journal of machine learning research}~{\bf 12},  2825 (2011).

\bibitem{Merker_Depth_from_Normals_2022}
Merker, E., ``{Depth from Normals},'' (2022).

\end{thebibliography}

\end{document}